\documentclass[epj,english]{webofc}
\usepackage[varg]{txfonts}   
\usepackage{comment}
\usepackage{textcomp}
\usepackage[usenames,svgnames,table]{xcolor} 

\newcommand{\degree}{\ensuremath{^\circ}}
\newcommand{\dee}{\ensuremath{\varDelta E-E}}
\woctitle{INPC 2013}
\begin{document}
\title{Measurement of light charged particles in the decay channels of medium-mass excited compound nuclei}
%
%

\author{S. Valdré\inst{1,2}\fnsep\thanks{\email{valdre@fi.infn.it}} \and  S. Barlini \inst{1} \and
G. Casini\inst{2} \and G. Pasquali\inst{1,2} \and S. Piantelli\inst{2} \and S. Carboni\inst{1,2} \and
M. Cinausero\inst{3} \and F. Gramegna\inst{3} \and T. Marchi\inst{3,4} \and
G. Baiocco\inst{5} \and L. Bardelli\inst{2} \and G. Benzoni\inst{6} \and M. Bini\inst{1,2} \and
N. Blasi\inst{6} \and A. Bracco\inst{6,7} \and S. Brambilla\inst{6} \and M. Bruno\inst{5} \and
F. Camera\inst{6,7} \and A. Corsi\inst{6,7} \and F. Crespi\inst{6,7} \and M. D'Agostino\inst{5} \and
M. Degerlier\inst{3} \and V. L. Kravchuk\inst{3} \and S. Leoni\inst{6,7} \and B. Million\inst{6,7} \and
D. Montanari\inst{4} \and L. Morelli\inst{5} \and A. Nannini\inst{2} \and R. Nicolini\inst{6,7} \and
G. Poggi\inst{1,2} \and G. Vannini\inst{5} \and O. Wieland\inst{6} \and
P. Bednarczyk\inst{8} \and M. Ciema\l{}a\inst{8} \and J. Dudek\inst{9} \and B. Fornal\inst{8} \and
M. Kmiecik\inst{8} \and A. Maj\inst{8} \and M. Matejska-Minda\inst{8} \and K. Mazurek\inst{8} \and
W. M\k{e}czy\'{n}ski\inst{8} \and S. Myalski\inst{8} \and J. Stycze\'{n}\inst{8} \and M. Zi\k{e}bli\'{n}ski\inst{8}
}

\institute{
	Dipartimento di Fisica, Università di Firenze, Sesto Fiorentino (FI), Italy
	\and
	INFN, Sezione di Firenze, Sesto Fiorentino (FI), Italy
	\and
	Laboratori Nazionali di Legnaro, Legnaro (PD), Italy
	\and
	Dipartimento di Fisica, Università di Padova, Padova, Italy
	\and
	Dipartimento di Fisica, Università di Bologna and INFN, Sezione di Bologna, Bologna, Italy
	\and
	INFN, Sezione di Milano, Milano, Italy
	\and
	Dipartimento di Fisica, Università di Milano, Milano, Italy
	\and
	Niewodnicza\'{n}ski Institute of Nuclear Physics, PAN, 31-342 Krakow, Poland
	\and
	Institut Pluridisciplinaire Hubert Curien, Strasbourg, France
}

\abstract{%
The $\mathrm{^{48}Ti}$ on $\mathrm{^{40}Ca}$ reactions have been studied at 300 and 600 MeV
focusing on the fusion-evaporation (FE) and fusion-fission (FF) exit channels.
Energy spectra and multiplicities of the emitted light charged particles
have been compared to Monte Carlo simulations based on the statistical model.

Indeed, in this mass region ($A\sim100$) models predict that shape transitions can occur at high spin values and
relatively scarce data exist in the literature about coincidence measurements between
evaporation residues and light charged particles. Signals of shape
transitions can be found in the variations
of the lineshape of high energy gamma rays emitted from the de-excitation of GDR states
gated on different region of angular momenta. For this purpose it is important
to keep under control the FE and FF processes, to regulate the
statistical model parameters and to control the onset of possible
pre-equilibrium emissions from 300 to 600 MeV bombarding energy.
}

\maketitle

\section{Introduction}
\label{sec:intro}
Macroscopic nuclear models~\cite{Mazurek11} predict a rapid
transition from oblate to prolate shape at high spin for
medium-mass nuclei; this transition can be observed thanks to the
low-fissility of these systems which can therefore sustain high spins with
limited fission probability. Variations of the GDR spectra can signal
this process but, since the effects are small, it is important an
accurate description of the decay of the excited compound nuclei (CN).
With this in mind and considering that not many data in this region exist from exclusive
measurements \cite{Sanders91,VonOertzen08,Charity10},
we performed an experimental campaign at the Tandem-ALPI complex of the Laboratori Nazionali di Legnaro.
$\mathrm{^{88}Mo}$ compoud nuclei have been studied, formed in the reactions of $\mathrm{^{48}Ti}$ on
$\mathrm{^{40}Ca}$ target (500 \textmu g/$\mathrm{cm^2}$) at 300 and 600 MeV bombarding energies.
Table~\ref{tab:parametri} reports some parameters for these reactions.

\begin{table}[h!]
\centering
\caption{
	Compound nucleus excitation energy, vanishing fission barrier spin and grazing angular momentum and estimated total reaction cross section
	reported for both bombarding energies.
}
\label{tab:parametri}
\begin{tabular}{ccccc}
\hline
$\boldsymbol{E}\;\mathbf{[MeV]}$ & $\boldsymbol{\varepsilon^*}\;\mathbf{\left[{MeV/u}\right] }$ & $\boldsymbol{l_0}\;(B_f=0)\;\mathbf{\left[{\hbar}\right]}$ & $\boldsymbol{l_{gr}}\;\mathbf{\left[{\hbar}\right]}$ & $\boldsymbol{\sigma_{R}}\;\mathbf{[barn]}$\\
\hline
300 & 1.4 & 64 & 88.4  & 1.7\\
600 & 3.0 & 64 & 149.2 & 2.5\\
\hline
\end{tabular}
\end{table}

The reactions have been analyzed focusing on the fusion-evaporation (FE) and fusion-fission (FF) exit channels,
which dominate in mass-symmetric collisions at moderate bombarding energies.
For excitation energies up to around 3 MeV/u the equilibrium of the internal degrees of freedom is reached
in shorter times compared to evaporation and fission time scales.
Thus it is reasonable to adopt, as a first approximation, the statistical model of compound nucleus to study its decay.
Experimental data are compared with events produced through Monte Carlo simulations
(performed with the well known Gemini++ code \cite{Charity10}) where complete fusion events are generated
with a triangular spin distribution for the CN up to $l_0$.
The output of the simulation is then filtered through
solid angle acceptance and particle energy detection thresholds
of the experimental apparatus.

\section{Experimental apparatus}
\label{sec:apparatus}


\begin{figure}
\centering
\includegraphics[width=8cm]{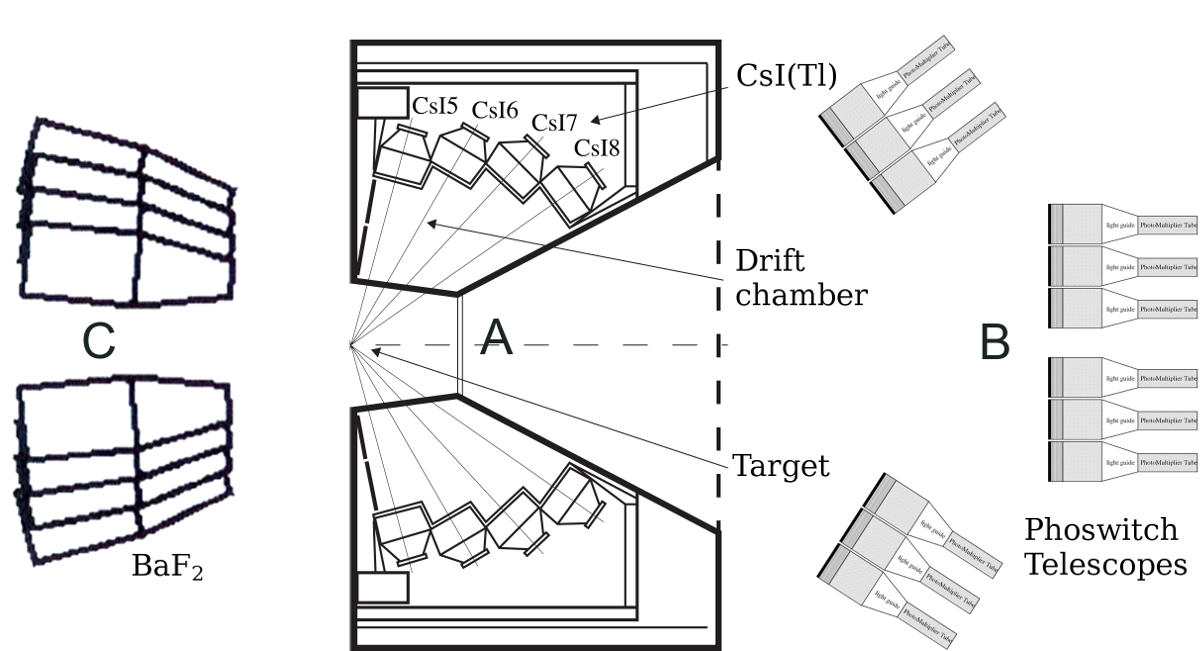}
\caption{The experimental setup was composed of the GARFIELD apparatus (A), a wall of Phoswich detectors (B)
and the HECTOR $\mathrm{BaF_2}$ crystals (C).}
\label{fig:garf}
\end{figure}

The experimental setup is shown in Figure~\ref{fig:garf}.
Phoswich telescopes \cite{Bini03} (made of two plastic scintillators and a CsI(Tl) crystal)
were used to select the exit channel (FE/FF) by detecting heavy residues or fragments in the polar range
from 6\degree\ to 12\degree , also allowing charge identification up to about $Z\sim 12$.
The forward chamber of the GARFIELD apparatus \cite{Gramegna97,Bruno13} was the main detector for the LCP emitted in the reaction.
It consists of 96 \dee\ telescopes where the $\varDelta E$ stage is a gas detector with moderate
multiplication, followed by a CsI(Tl) crystal 4 cm thick.
The 96 telescopes are organized in four rings (referred as 8, 7, 6, 5), from 29\degree\ to 85\degree\ (polar range) and 24 azimuthal sectors,
thus covering a big part of the forward emisphere ($\sim 1.5\pi$ sr).
At backward angles (>90\degree) the HECTOR array made of eight big $\mathrm{BaF_2}$ crystals
was used to measure hard gamma rays associated to GDR.

The $E$ vs $tof$ technique was used to detect the heavy fragments (and hence also ER) in the Phoswich detectors.
Via \dee\ correlations the Phoswich also allowed charge identification for intermediate mass fragments (IMF) and
mass determination for light charged particles (LCP).

\begin{figure}[htbp]
\centering
\includegraphics[width=0.46\textwidth]{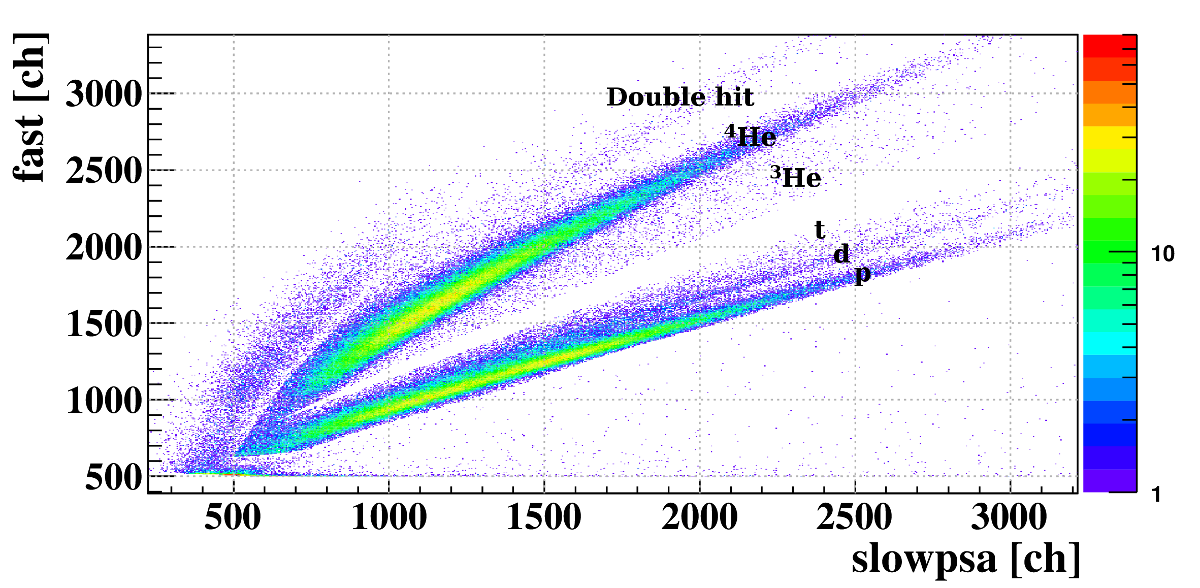}
\hspace*{2mm}
\includegraphics[width=0.43\textwidth]{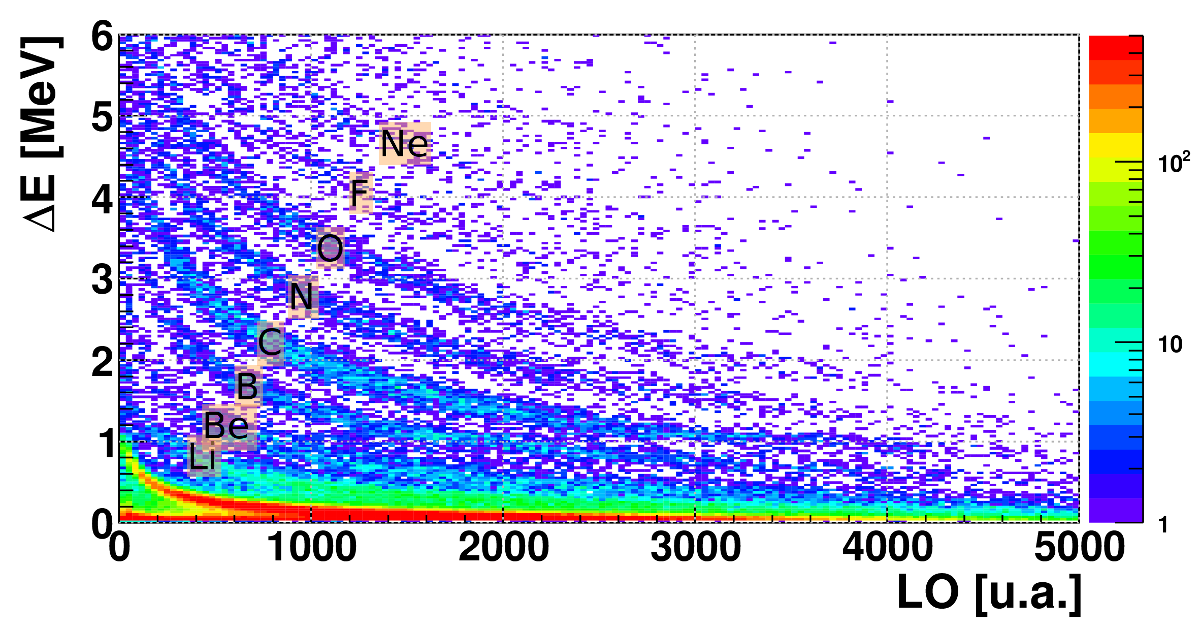}
\caption{Particle identification in GARFIELD. Isotopes of light charged particles are discriminated via $fast-slow$ technique (left).
Instead \dee\ (right) is used for IMF charge identification.}
\label{fig:garfid}
\end{figure}

LCP in GARFIELD are identified through the usual $fast-slow$ correlation in CsI(Tl) crystals
as shown in Figure~\ref{fig:garfid} (left panel). Heavier fragments are identified with the \dee\ method
using the gas stage (right panel of Figure~\ref{fig:garfid}).

\section{Data analysis}
To obtain ``$4\pi$'' particle multiplicities we used simulated data to correct for the efficiency of the apparatus.
In particular, for each particle species, the multiplicity measured in a given set of detectors has been divided
by the efficiency of the same set of detectors, evaluated by the simulations.
The LCP multiplicities (obtained from Phoswich data only) in coincidence with an ER are shown in Table~\ref{tab:multphos}
for the two measured energies and they are compared to statistical model predictions.

\begin{table}[h!]
\centering
\caption{
	Number of LCP emitted per ER. Yields are obtained from Phoswich detectors only and they are corrected for the efficiency.
	Not enough statistics has  been measured for tritons which are not reported.
}
\label{tab:multphos}
\begin{tabular}{ccccccc}
\hline
 & \multicolumn{3}{c}{\bfseries 300 MeV} & \multicolumn{3}{c}{\bfseries 600 MeV} \\
 & $\mathbf{m_p}$ & $\mathbf{m_d}$ & $\mathbf{m_\alpha}$ & $\mathbf{m_p}$ & $\mathbf{m_d}$ & $\mathbf{m_\alpha}$\\
\hline
\bfseries EXP & $4.0 \pm 0.8$ & $0.18 \pm 0.07$ & $2.9 \pm 0.6$ & $5.5 \pm 1.1$ & $0.7 \pm 0.3$ & $3.8 \pm 0.8$\\
\bfseries MC & 3.5 & 0.08 & 1.1 & 5.9 & 0.53 & 1.8\\
\hline
\end{tabular}
\end{table}

As expected, the measured (and predicted) multiplicities at 600 MeV are larger than at 300 MeV.
Protons appear to be reasonably reproduced by the statistical model, run with default values of input parameters \cite{Charity10}.
Instead, at both energies data show an excess of $\alpha$ particles with respect to Gemini++ predictions.
At 300 MeV also deutons are more abudantly emitted with respect to calculations.
The evaluation of LCP multiplicities also using GARFIELD data is in progress.

\begin{table}[h!]
\centering
\caption{LCP yield ratios from GARFIELD and Phoswich data in the FE channel.}
\label{tab:ratios}
\begin{tabular}{ccccc}
\hline
& \multicolumn{2}{c}{\bfseries 300 MeV} & \multicolumn{2}{c}{\bfseries 600 MeV} \\
& \bfseries p/d & \bfseries p/$\boldsymbol{\alpha}$ & \bfseries p/d & \bfseries p/$\boldsymbol{\alpha}$\\
\hline
\bfseries EXP from GARFIELD & $29 \pm 12$ & $2.9 \pm 0.6$  & $9.4 \pm 3.8$ & $2.4 \pm 0.5$\\
\bfseries EXP from Phoswich & $22 \pm 9$  & $1.4 \pm 0.3$  & $7.8 \pm 3.1$ & $1.4 \pm 0.3$\\
\bfseries MC & 42.3 & 3.30 & 11.3 & 3.39\\
\hline
\end{tabular}
\end{table}

The p/d and p/$\alpha$ ratios obtained from the Phoswiches and GARFIELD are separately reported in Table~\ref{tab:ratios}.
From Table~\ref{tab:multphos} and Table~\ref{tab:ratios} one can conclude that more $\alpha$-particles
are emitted than predicted by Gemini++ run with standard parameters, and that the excess increases
at forward angles (Phoswich region).

\begin{figure}[htbp]
\centering
\includegraphics[width=0.48\textwidth]{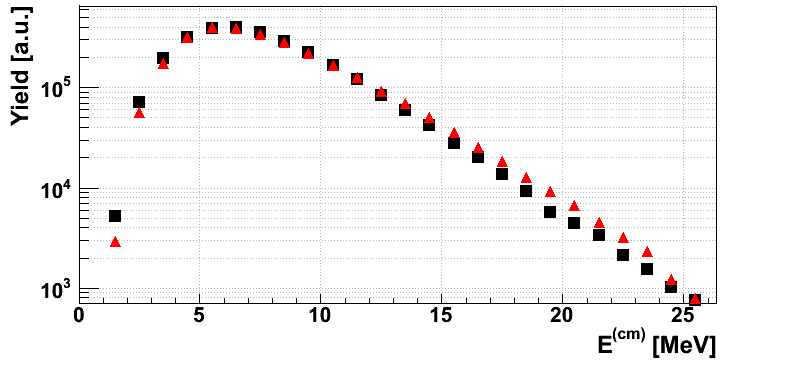}
\includegraphics[width=0.48\textwidth]{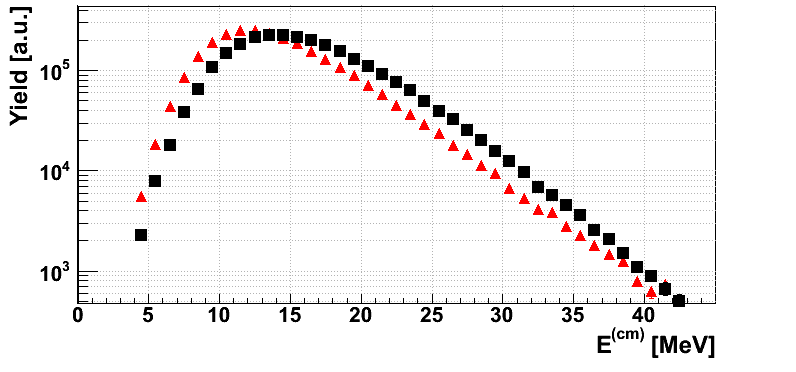}
\caption{
	Center of mass energy spectra of LCP detected in GARFIELD between $\theta=29.5\degree$ and $\theta=40.0\degree$ (\textit{ring 8})
	in the fusion-evaporation channel at 300 MeV. Proton spectrum is on the left while $\alpha$-particle spectrum is on the right.
	Black squares are experimental points, while red triangles are simulated data.
	The yields are normalized to the integral for each panel.
}
\label{fig:garfspc}
\end{figure}

The shape of energy spectra (Figure~\ref{fig:garfspc}) seems to confirm the pure statistical nature of proton emission while
some disagreement is found for $\alpha$ particles. In particular, $\alpha$ particles appear to be on average more energetic
than predicted and this happens also for the FF channel.

\section{Conclusions and remarks}
The decay of $\mathrm{^{88}Mo}$ at 1.4 and 3.0 MeV/u excitation energy is under study.
Proton emission in the FE channel is well reproduced by the Gemini++ statistical code
run with standard input values.
An excess of $\alpha$ particles has been found at both bombarding
energies also associated to a slightly different CM-energy spectrum
with respect to the predicted one.
Work is in progress to better quantify this excess and to understand if it could be accounted
for by suitable variations of model parameters, in particular to better describe the shape deformation
at high spin as pointed out in \cite{Fornal90}, or it comes for other sources.
It is also in progress the study of the fission channel,
whose description is important for shape transitions as it is associated to the
highest spins where such shape changes are expected.

\bibliography{biblio}

\end{document}